\documentclass[prl,nofootinbib,twocolumn,
unsortedaddress]{revtex4}
\usepackage{xr}
\usepackage[colorlinks=true,linkcolor=black]{hyperref}
\usepackage{graphicx}
\usepackage[utf8]{inputenc}
\usepackage{amsmath}
\usepackage{amssymb}
\usepackage{amsthm}
\usepackage{amsfonts}
\usepackage{booktabs}
\usepackage{bbm}
\usepackage[capitalise]{cleveref}
\usepackage{dsfont}
\usepackage[dvipsnames]{xcolor}
\usepackage{esint}
\usepackage{slashed}
\usepackage{siunitx}
\usepackage{tabularx}

\begin{document}

\title{Comment on ``Strong Quantum Darwinism and Strong Independence
are Equivalent to Spectrum Broadcast Structure''}

\author{Alexandre Feller$^1$}
\author{Benjamin Roussel$^{1}$}
\author{Irénée Frérot$^{2,3}$}
\author{Pascal Degiovanni$^4$}

\affiliation{(1) European Space Agency - Advanced Concepts Team,
ESTEC, Keplerlaan 1,
2201 AZ Noordwijk,
The Netherlands.}

\affiliation{(2) ICFO-Institut de Ciencies Fotoniques, The Barcelona Institute of Science and Technology, Av. Carl Friedrich Gauss 3, 08860 Barcelona, Spain
}
\affiliation{(3) Max-Planck-Institut f{\"u}r Quantenoptik, D-85748 Garching, Germany}

\affiliation{(4) Univ Lyon, Ens de Lyon, Universit\'e Claude Bernard 
Lyon 1, CNRS, 
Laboratoire de Physique, F-69342 Lyon, France}

\date{\today}
\begin{abstract}
In a recent Letter [Phys.~Rev.~Lett.~122, 010403 (2019)], an equivalence is proposed between the so-called Spectrum Broadcast Structure for a system-multienvironment quantum state, and the conjunction of two information-theory notions: (a) Strong Quantum Darwinism; and (b) Strong Independence. Here, we show that the mathematical formulation of condition (b) by the authors (namely, the pairwise independence of the fragments of the environment, conditioned on the system), is necessary but not sufficient to ensure the equivalence. We propose a simple counter-example, together with a strengthened formulation of condition (b), ensuring the equivalence proposed by the authors. 

\end{abstract}
\maketitle

In their paper \cite{Castro-2019}, the authors introduce the notions of: (a) 
``strong quantum Darwinism''; and (b) ``strong independence''. 
As the main result of the paper, it is proposed that, taken together,
conditions (a) and (b) are equivalent to the so-called
``spectrum broadcast structure'' (SBS) for the system--environment
quantum state $\rho_{{\cal S}{\cal E}}$ (throughout this Comment, we follow the notations and definitions used in the paper). By definition, if $\rho_{{\cal S}{\cal E}}$
has a SBS, the state of ${\cal E}={\cal E}_1\cdots {\cal E}_F$ relative to
${\cal S}$ is fully factorized, and has \textit{no correlations whatsoever}; and the authors aimed at capturing this fully-factorized structure by an information-theory criterion of ``strong independence''.
In this Comment, we show that condition (b), while ensuring
\textit{pairwise} independence of the fragments ${\cal E}_i$ when
conditioning on the system (a necessary condition for having the SBS), does not rule out the existence of higher-order correlations, and as such is insufficient to imply a SBS. We first construct a
(completely classical) counter-example of a state with no SBS yet satisfying
conditions (a) and (b). We then propose a stronger condition (b'), ensuring
the equivalence of conditions (a) and (b') with SBS. At a conceptual level, the main conclusion of the paper, namely the equivalence between, on the one hand, the SBS, and on the other, the conjunction of ``strong quantum Darwinism'' (as defined by the authors) and ``strong independence'' (as now defined by (b')), remains therefore unaltered. 

\noindent \textit{Counter-example.} 
We consider a qubit system ${\cal S}: \{|0\rangle, |1\rangle\}$, and the
environment ${\cal E} = {\cal E}_1{\cal E}_2{\cal E}_3$, with three
qutrits fragments 
${\cal E}_i: \{|0_i\rangle, |1_i\rangle, |2_i \rangle\}$ ($i\in \{1,2,3\}$).
We define the projectors $\Pi_a^{({\cal S})} = |a\rangle \langle a|$ and
$\Pi_a^{(i)} = |a_i\rangle \langle a_i|$. Finally, we define the projectors
$\Pi_{abc} = \Pi_a^{(1)} \otimes \Pi_b^{(2)} \otimes \Pi_c^{(3)}$. Our
counter-example is the state ($0\le p \le 1$):
\begin{equation}
	\rho_{{\cal S}{\cal E}} = (1-p) ~\Pi_0^{({\cal S})} \otimes \rho_{{\cal E}}^{(0)} +
		p~ \Pi_1^{({\cal S})} \otimes \rho_{{\cal E}}^{(1)} ~,
\end{equation}
with the relative states:
\begin{eqnarray}
	\rho_{{\cal E}}^{(0)} &=& \Pi_{000} \\
	\rho_{{\cal E}}^{(1)} &=& \frac{1}{4}[\Pi_{111} + \Pi_{122} + \Pi_{212} + \Pi_{221}] \label{eq_rel_123}\\
	\rho_{{\cal E}_i {\cal E}_j}^{(1)} &=& \frac{1}{2}[\Pi_{1}^{(i)} + \Pi_2^{(i)}] \otimes \frac{1}{2}[\Pi_{1}^{(j)} + \Pi_2^{(j)}] \label{eq_rel_ij}\\
	\rho_{{\cal E}_i}^{(1)} &=& \frac{1}{2}[\Pi_1^{(i)} + \Pi_2^{(i)}]
\end{eqnarray}
As the relative states $\rho_{{\cal E}_i}^{(0)} = \Pi_0^{(i)}$ and 
$\rho_{{\cal E}_i}^{(1)} = [\Pi_1^{(i)} + \Pi_2^{(i)}] / 2$ have disjoint
supports, it is clear that each fragment has full access to the system's state.
Therefore: 
$I({\cal S}:{\cal E}_i) = I_{\rm acc}({\cal S}:{\cal E}_i) 
= \chi({\cal S}^\Pi: {\cal E}_i) = H({\cal S})$ [condition (a)].
The state also fulfills condition (b) (pairwise-independence conditioned
on ${\cal S}$). Indeed, the relative states 
$\rho_{{\cal E}_i {\cal E}_j}^{(0)} = \Pi_0^{(i)} \otimes \Pi_0^{(j)}$ and
$\rho_{{\cal E}_i {\cal E}_j}^{(1)}$ [Eq.~\eqref{eq_rel_ij}] are product states,
which implies that $I({\cal E}_i:{\cal E}_j|{\cal S}) = 0$ (as can be 
checked by direct computation on $\rho_{{\cal S}{\cal E}_i{\cal E}_j}$). Yet, the state 
$\rho_{{\cal S}{\cal E}_1{\cal E}_2{\cal E}_3}$ does \textit{not} admit
a SBS. Indeed, the relative state
$\rho_{{\cal E}}^{(1)}$ [Eq.~\eqref{eq_rel_123}] does \textit{not}
factorize (while pairs of fragments are uncorrelated, the parity of the
total number of fragments in state $|1_i\rangle$ is odd, yielding genuine
3-partite correlations). This concludes our proof that the conditions
(a) and (b) are insufficient to imply a SBS.

\noindent \textit{Proposed correct formulation of the theorem.}
The above counter-example suggests to introduce a stronger condition
(b') ensuring the complete factorization of the fragments when
conditioning on the system, which is a defining property of the SBS.
Noticing that a multipartite state is factorized, 
$\rho_{{\cal E}_1, \dots {\cal E}_F} = \otimes_{i=1}^F \rho_{{\cal E}_i}$,
iff the multipartite mutual information vanishes\cite{Yang-2009}:
$I({\cal E}_1, \dots {\cal E}_F) := 
\sum_{i=1}^F H(\rho_{{\cal E}_i})  - H(\rho_{{\cal E}_1, \dots {\cal E}_F}) = 0$,
we are led to propose the following:\\
\noindent\textbf{Theorem.} \textit{
The state $\rho_{{\cal S}{\cal E}}$ has the SBS iff:
\begin{itemize}
\item[(a1)] $I({\cal S}:{\cal E}) = \chi({\cal S}^\Pi,{\cal E})$ (classical--quantum state),
\item[(a2)] $I_{\rm acc}({\cal S}:{\cal E}_i) = H({\cal S})$ for all $i$
	(the information about ${\cal S}$ can be fully recovered by measuring any fragment),
\item[(b')] $I({\cal E}_1, \cdots {\cal E}_F | {\cal S}) = 0$ (totally-factorized 
relative states).
\end{itemize}
}
\begin{proof}
 If the state has the SBS, it is clear that conditions (a1), (a2) and (b') are 
 fulfilled. Conversely, condition (a1) (vanishing discord) implies
 that the system--environment state is of the form 
 $\rho_{{\cal S}{\cal E}} = \sum_s p_s \Pi_s^{({\cal S})} 
 \otimes \rho_{\cal E}^{(s)}$ with $\{\Pi_s^{({\cal S})}\}$
 forming mutually orthogonal projectors for the system. Condition (b')
then implies that the relative states factorize: 
$\rho_{\cal E}^{(s)} = \otimes_{i=1}^F \rho_{{\cal E}_i}^{(s)}$.
Finally, condition (a2) implies that for each $i$, the states 
$\rho_{{\cal E}_i}^{(s)}$ are pairwise orthogonal. 
\end{proof}

\acknowledgments{We thank Thao P.~Le and Alexandra Olaya-Castro for
constructive feedback on our manuscript. 
This work has been supported by the European Space Agency (Ariadna
study 1912-01). IF acknowledges support from the Government of Spain
(FIS2020-TRANQI and Severo Ochoa CEX2019-000910-S), Fundaci{\'o} Cellex
and Fundaci{\'o} Mir-Puig through an ICFO-MPQ Postdoctoral Fellowship,
Generalitat de Catalunya (CERCA, AGAUR SGR 1381 and QuantumCAT).}

\bibliographystyle{amsplain}
\bibliography{biblio}

\end{document}